\documentclass[a4paper,11pt]{article}
\usepackage{pos}

\title{Theory of the (Heavy-Quark) Exotic Hadrons:\\A Primer for the Flavor Community}\ShortTitle{Theory of the (Heavy-Quark) Exotic Hadrons}

\author*[a]{Richard F. Lebed}

\affiliation[a]{Arizona State University,\\
Department of Physics, Tempe, AZ 82587-1504 \ USA}

\emailAdd{richard.lebed@asu.edu}

\abstract{Scores of exotic hadrons, particularly tetraquarks and pentaquarks in the heavy-quark sector, have been observed in the past 20 years, and more continue to be discovered to this day.  Unlike me\-sons and baryons, such exotics are not mandated to exist under our current understanding of QCD, meaning that each new discovery presents fresh insights into the expansive possibilities of strong-interaction physics.  Even the basic architecture of these multiquark states remains an open question.  Here we discuss the merits and deficiencies of their best-known dynamical descriptions, recognizing that the eventual universal model for exotics will almost certainly require the synthesis of more than one fundamental paradigm.}

\FullConference{21st Conference on Flavor Physics and CP Violation (FPCP 2023)\\
 29 May - 2 June 2023\\
 Lyon, France\\}

\begin{document}
\maketitle

\section{Introduction}

While this presentation is general in scope, it is designed especially for researchers in flavor physics, who specialize in studies of what makes the dynamics of different flavor sectors unique: $C\!P$ violation, CKM elements, rare hadronic decays, lepton-flavor violation, and so on.  The core message is that a series of remarkable, unexpected discoveries has occurred in the past 20 years, particularly in the heavy-quark ($c$ and $b$) sector, revealing the existence of new types of multiquark hadrons.  As will be discussed, the era of these discoveries is far from over, and even simple questions such as the basic structure of these states remains unresolved.

To begin this overview, we remind the reader of one simple, familiar fact about the SU(3)$_{\rm color}$ gauge symmetry of QCD\@.  Color confinement requires hadrons to be color neutral, and the two most elementary ways to satisfy this constraint with quarks that transform under the fundamental color-{\bf 3} of  SU(3)$_{\rm color}$ are as $q\bar q$ mesons or $qqq$ baryons.  Color charge is so named because, like colors of light, it provides two distinct ways to make color-neutral states: by combining a color and its opposite ({\it e.g.}, red + cyan) or by combining equal amounts of the three primary colors (red + blue + green).  From this fundamental physics, it is not hard to see that {\em mesons and baryons are necessary}.  Suppose you create a universe containing color-{\bf 3} quarks (and color-$\bar{\bf 3}$ antiquarks).  Start with the universe in a hot, dense state, and demand only the existence of color confinement and the suppression of baryon-number violation at low energies.  As the universe begins to cool, anyplace that contains a mixture of $q$ and $\bar q$ will naturally experience the condensation of mesons, while anyplace that contains an excess of $q$ ($\bar q$) will naturally experience the condensation of baryons (antibaryons).  Indeed, this sketch embodies features of conventional early-universe models.

On the other hand, any hadrons more complicated than mesons or baryons that might arise in this way can be expected to fall apart into the more elementary hadron classes, or might not even exist as bound or resonant states at all.  In this sense, {\em no other classes of hadrons are necessary}.  Nevertheless, mesons and baryons are not the only possibilities; any other class of hadrons (collectively called {\it exotics}) that can be observed tells us a lot about the nature of QCD\@.  The general SU(3)$_{\rm color}$ rule for forming all possible color-neutral states is simple: Allowed states satisfy (\# of $q$) – (\# of $\bar q$) = 0 mod 3, plus any number of gluons $g$ except one in isolation.  Including the possibility of $g$ to provide contributions to the valence quantum numbers of the state, one can then exhaustively list all the exotic hadron types:
\begin{itemize}
\item $gg$, $ggg$, \ldots (glueball)
\item $q\bar q g$, $q\bar q gg$, \ldots (hybrid meson)
\item $q\bar q q\bar q$, $q\bar q q\bar q q\bar q$, \ldots (tetraquark, hexaquark, \ldots)
\item $qqqq\bar q$, $qqqqqqq\bar q$, \ldots (pentaquark, octoquark, \ldots)
\item $qqqqqq$, \ldots (dibaryon, \ldots)
\end{itemize}
Such a catalog of potential states is hardly new; even Gell-Mann~\cite{Gell-Mann:1964ewy} and Zweig~\cite{Zweig:1964ruk,Zweig:1964jf} in their foundational 1964 quark-model papers noted the multiquark possibilities.

A quick historical summary serves to put the discovery of exotics into context.  For decades, hadron spectroscopy was the core activity of high-energy physics (see, {\it e.g.}, Ref.~\cite{Pais:1986nu}).  Starting immediately after World War~II with the 1947 discovery of $\pi^{\pm}$, $K^{\pm}$, $K^0$, the period 1950--1965 saw an explosion of discoveries (the so-called ``hadron zoo''), which were gradually clarified through the development of strangeness, the Eightfold Way, the quark model, and color charge.  1974 saw the discovery of charmonium, which went hand-in-hand with experimental evidence for asymptotic freedom and QCD as the gauge theory of strong interactions.  In 1977 bottomonium was first observed, providing the $3^{\rm rd}$ generation of quarks necessary to allow for the CKM theory of $C\!P$ violation.  Finally, 1983 saw two important experimental developments that produced very different effects; one was the first full reconstruction of $B$ meson decays by CLEO~\cite{CLEO:1983mma}, which opened the way to studying hadron spectroscopy in virtually every flavor sector.  The other was the discovery of the $W$ and $Z$ bosons at CERN, which inspired a large number of researchers in high-energy physics to focus upon predictions for the top quark, the Higgs boson, and beyond-Standard Model theories.  Afterwards, hadron spectroscopy became much less central to the main effort of particle physics, being viewed by much of the field as merely the exercise of filling out quark-model multiplets.

\section{Hadron Renaissance}

A remarkable accidental discovery in 2003 initiated a shift of attitudes regarding the predictability of hadron physics.  The Belle Collaboration found evidence~\cite{Choi:2003ue} for a narrow new particle at 3872~MeV ($\approx 4 m_p$), decaying to $J/\psi \pi^+ \pi^-$, that behaves very unlike a pure $c\bar c$ state.  This $X(3872)$ [``$X$'' simply indicating ``unknown''], is a ``charmoniumlike'' state (meaning that all of its known decays contain a $c\bar c$ pair), and is almost certainly a hadron of valence quark content $c\bar c q \bar q$.  The measured quantum numbers $J^{PC} = 1^{++}$ lead to its Particle Data Group (PDG)~\cite{ParticleDataGroup:2022pth} label $\chi_{c1}(3872)$.  As an important reminder, the primary physics goal of Belle was the search for $C\!P$ violation in the $B$ system which, prior to its run, had only been seen in neutral $K$ decays.

\begin{figure}
    \centering
    \vspace{-17ex}
    \hspace{-3em}
    \includegraphics[scale=0.75]{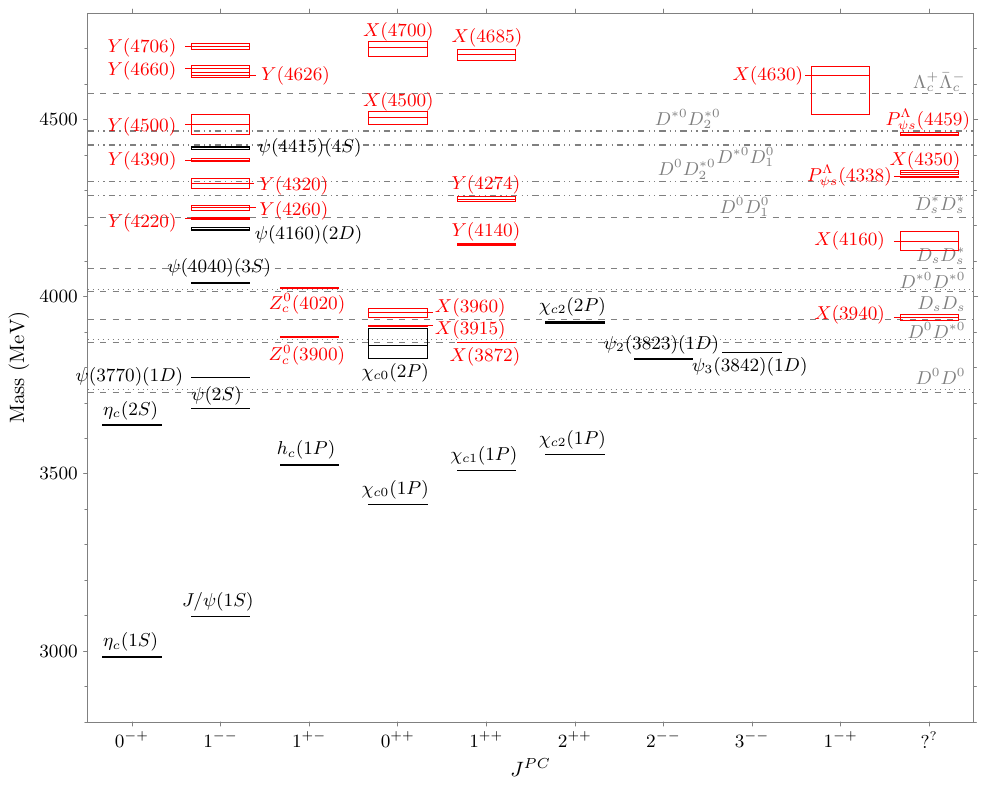}
    \vspace{-47ex}
    \caption{Level diagram for the neutral charmoniumlike system in June, 2023\@.  Conventional $c\bar c$ states are presented in black, exotic candidates in red.  The $c\bar c c\bar c$ candidates like $X(6900)$ would appear far off the top of this diagram.  Boxes indicate uncertainties on mass measurements.  Dashed and dotted lines denote di-hadron mass thresholds.}
    \label{fig:NeutralccLevel}
\end{figure}

Fast-forwarding 20 years, the situation has turned out to be much richer than anyone could have anticipated.  First consider just the system of hidden-charm hadrons, represented in any particle-physics textbook by a level diagram exhibiting the predicted and known conventional charmonium states.  The current equivalent diagram (Fig.~\ref{fig:NeutralccLevel}) clearly features far more states than expected.

One conspicuous feature of Fig.~\ref{fig:NeutralccLevel} is the proximity of several of the states to di-hadron thresholds.  The most prominent example is actually $X(3872)$ itself: $m_{X(3872)} - m_{D^0} - m_{D^{*0}} = -40 \pm 90$~keV~\cite{ParticleDataGroup:2022pth}.  In comparison, the deuteron, which is already considered to be a rather weakly bound state, has $m_d - m_p - m_n = -2.2452(2)$~MeV\@.  However, many of these states are not particularly close to any such thresholds; prominent examples include the $1^{--}$ states, generically called ``$Y$'' by the community and denoted $\psi$ by the PDG\@.  The lightest and best-studied example is $Y(4220)$, which has been seen in multiple decay channels.

Similar conclusions hold for the charged hidden-charm system, which includes observed nonstrange ($Z_c$) and strange ($Z_{cs}$) tetraquark candidates [the lightest example being $Z_c^+(3900)$], as well as pentaquark ($P_c$) candidates.  The same labels are used for their neutral isospin partners.  But it is also worth noting that a charged charmoniumlike state like $Z^+_c$ is manifestly exotic, since its $c\bar c$ component is electrically neutral, so its minimal quark content is $c\bar c u \bar d$.

So then, what is the result of the heavy-quark exotics census as of June, 2023?  By my count, the experimental collaborations have reported {\bf 64} observed heavy-quark exotics, both tetraquarks and pentaquarks.  Some of them may turn out to not to be confirmed by other experiments, and some might turn out to be conventional quarkonium states, but even if true, the tally still remains impressive.  Of these 64, {\bf 49} are in the hidden-charm sector (both electrically neutral and charged, including states with open strangeness).  {\bf 5} occur in the (much less explored) hidden-bottom sector, {\bf 4} have a single $c$ quark (and an $s$, a $u$, and a $d$), {\bf 1} has a single $b$ quark (and an $s$, a $u$, and a $d$), {\bf 4} have valence-quark content $c\bar c c\bar c$, and {\bf 1} has two $c$ quarks.  If one simply examines every flavor sector and makes any reasonable assumptions about the existence of orbitally and radially excited states, then a naive count estimates well over 100 more exotics awaiting discovery.

And not all exotic candidates have heavy quarks.  Some of the more conspicuous examples include the isotriplet $\pi_1 (1600)$~\cite{E852:1997gvf} discovered by the Brookhaven E852 Collaboration, which is believed to be a hybrid meson because its $J^{PC} = 1^{-+}$ quantum numbers are not accessible to $q\bar q$ states.  The scalar-isoscalar $f_0 (1710)$ is expected to contain a sizeable glueball component because the quark model predicts one fewer unflavored $0^{++}$ states than are seen experimentally, and of them, $f_0 (1710)$ shows up most prominently in $J/\psi$ decays (which is a glue-rich environment because its decays require $c\bar c$ annihilation).  The state $\phi (2170)$ (so named because it is $J^{PC} = 1^{--}$ and is dominated by decay channels containing $s\bar s$) has peculiar decay patterns that may indicate it to be an $s\bar s g$ hybrid or the $s\bar s q\bar q$ tetraquark analogue to the $c\bar c q\bar q$ $1^{--}$ state $Y(4220)$.  Several other examples are known that exhibit hints of exotic behavior.

\section{Models for the Exotics}

Even after 20 years of multiple experimental advances, the definitive prescription for describing the internal structure of the exotics remains unknown.  Theory has nevertheless remained far from silent in investigating different scenarios for explaining their nature.  In essence, these pictures differ in the way that the quarks are organized within the full state.  Here we survey the best-explored proposals for tetraquarks, with the understanding that each possible structure has a pentaquark analogue.  Given a state consisting of 4 valence quarks $\bar Q_1, Q_2, \bar q_3, q_4$ (where $Q$ indicates a heavy quark), the dynamical paradigms to be summarized in the remainder of this presentation are:
\begin{itemize}

\item Hybrid mesons ($\bar Q_1 Q_2 g$)

\item Hadronic molecules [($\bar Q_1 q_4$)($Q_2 \bar q_3$)]

\item Diquark-antidiquark states [($\bar Q_1 \bar q_3$)($Q_2 q_4$)]

\item Hadroquarkonium [($\bar Q_1 Q_2$)($q_3 q_4$)]

\item Threshold/Rescattering/Cusp effects [interaction induced by the ($\bar Q_1 q_4$)($Q_2 \bar q_3$) threshold]

\end{itemize}
In addition, a few other techniques have been explored over the years, such as quark potential models, bag models, string-junction models, holography, and QCD sum-rule studies.  But the fact that no single approach has provided a universal description for all of the known states is exceptionally telling.  To quote the Snowmass~2021 white paper on multiquark hadrons~\cite{Brambilla:2022ura},
\begin{quote}
``Each of the interpretations provides a natural explanation of parts of the data, but neither explains all of the data. It is quite possible that both kinds of structures appear in Nature.  It may also be the case that certain states are superpositions of the compact and molecular configurations.''
\end{quote}

\noindent The concept of requiring quantum-mechanical superpositions of configurations to describe physical states is not unknown in hadron physics, but historically in such cases there has always been one overwhelmingly dominant component.  The properties of the system of exotics suggest that one must allow for a broader perspective.

Generally speaking, though, how can it be so hard to figure out what the exotics are?  When $J/\psi$ was discovered on November 11, 1974~\cite{E598:1974sol,SLAC-SP-017:1974ind}, it was immediately recognized to be essentially a bound state of a new, heavy charm quark and its antiquark, $c\bar c$, interacting through a scalar potential $V(r)$ created by innumerable gluons and sea $q\bar q$ pairs, whose collective effect in the lowest-mass states just supplies trivial quantum numbers, $J^{PC} = 0^{++}$.  Quarks much heavier than the QCD scale $\Lambda_{\rm QCD}$, act as discernable, nonrelativistic entities within the hadron.  Thus, quarkonium can be treated as a two-body problem in a simple scalar potential $V(r)$, and once $m_c$ also given, the Schr\"{o}dinger equation predicts the entire spectrum ({\it e.g.}, the Cornell potential~\cite{Eichten:1978tg,Eichten:1979ms}).  The states indicated by the black lines in Fig.~\ref{fig:NeutralccLevel} are predicted to occur right where they are observed.

The multiquark case is more complicated, primarily due to the presence of more dynamical degrees of freedom and no obvious principle to decide which ones are most significant for describing the state.  But even in such cases, having heavy quarks is helpful.  All common theoretical pictures for exotics rely in some way upon the fact that $\Lambda_{\rm QCD} / m_{c,b} \ll 1$.  Specific consequences of this limit especially significant to particular techniques are:
\begin{itemize}

\item $m_Q$ is nonrelativistic in the state, and has a small kinetic energy ({\em potential models, lattice simulations})

\item The heavy-quark Compton wavelength $\hbar / m_Q c$ is smaller than the full hadronic size, so that heavy quarks nucleate ``discernable'' localized clusters within the state ({\em hybrids, molecular models, diquark models, hadrocharmonium})

\item The scale $m_Q$ is heavy enough to belong to the asymptotic-freedom region of QCD, allowing for an operator expansion in powers of $1/m_Q$ ({\em heavy-quark spin symmetry, Born-Oppenheimer approximation, QCD sum rules})

\item Two-hadron thresholds for large $m_Q$ are spaced further apart than those in the tangled forest of states in the range 1--2~GeV (which is a major reason that light-quark exotics are hard to identify); such hadrons tend to possess fewer decay modes (and hence are narrower); the exotics also exhibit anomalous modes [{\it e.g.}, the isospin-violating channel $X(3872) \to J/\psi \rho$]

\end{itemize}

\subsection{Hybrid Mesons}

A hybrid is defined as any meson in which the valence quarks (including their relative orbital angular momentum) are insufficient to describe the state's overall quantum numbers.  More precisely, in a hybrid the quarks interact in a potential that is electrically neutral and isoscalar, but nonetheless carries nontrivial quantum numbers $J^{PC} \neq 0^{++}$ (``excited glue'').  In the simplest models for hybrids, the light degrees of freedom (d.o.f.) are subsumed into a ``constituent gluon'' quasiparticle; however, more sophisticated treatments consider a near-static $c\bar c$ (or other $\bar Q_1 Q_2$) quark pair connected by a {\em color flux tube\/} that carries nontrivial $J^{PC}$.  In this case, the light d.o.f.\@ generate {\em Born-Oppenheimer approximation\/} potentials with which the $c\bar c$ pair interacts.

Such systems have been studied in lattice-QCD simulations multiple times over the decades, stretching as far back as 1983~\cite{Griffiths:1983ah}.  All lattice simulations agree on ordering of the lowest multiplets in mass: In particular, the first nontrivial multiplet consists of states with $J^{PC} = (0,1,2)^{-+}, 1^{--}$.  Of these states, $1^{-+}$ is not allowed for $\bar Q_1 Q_2$ mesons, and hence is ``exotic'' not only in valence content, but in quantum numbers as well.  The $\pi_1 (1600)$ mentioned above is such a state, but with light valence quarks.

Modern lattice simulations also predict hybrid states with non-exotic quantum numbers, and their calculated mass eigenvalues can be compared to those of experimentally known states.  As an example, Ref.~\cite{Liu:2012ze} predicts a $1^{--}$ state at $4285 \pm 14$~MeV, right atop the lowest $Y$ states.  However, if a state like $Y(4220)$ is a hybrid, then why is it observed to transition prominently~\cite{BESIII:2020pov} to the charged, and hence non-hybrid, $Z^+_c (3900)$?  One sees that not only spectroscopy, but also decay patterns, can be important for discerning clues about the substructure of exotics.  For example, the decays of hybrids should obey distinctive selection rules ({\it e.g.}, no decays to two $S$-wave mesons~\cite{Page:1996rj}) that can help distinguish them from other structures.

\subsection{Hadronic Molecules}

Owing to the usual QCD rules for forming color singlets, one can construct multiquark states of any number of quarks $>3$, but the color structure of every one of them can be decomposed into that of a combination of (color-singlet) mesons and baryons: $(q\bar q)(q\bar q)$ for tetraquarks, $(qqq)(q\bar q)$ for pentaquarks, $(q\bar q)(q\bar q)(q\bar q)$ or $(qqq)(\bar q \bar q \bar q)$ for hexaquarks, $(qqq)(q\bar q)(q\bar q)$ for heptaquarks, {\it etc.}  It is therefore completely natural to gravitate toward the idea that any multiquark exotic is simply a molecule of mesons and/or baryons.  After all, atomic nuclei have successfully been modeled as nucleon bound states for decades. 

The most straightforward molecular paradigm is again modeled on an idea first proposed for atomic nuclei: that of hadrons bound by meson (dominantly $\pi$) Yukawa exchange.  The prototype for this picture is the $pn$ deuteron, which we have noted has a binding energy of 2.2~MeV\@.  In fact, the naive expectation for its binding energy is rather larger: $\Lambda_{\rm QCD}^2 / 2\mu \sim O(10~{\rm MeV})$, and so even in this most archetypical case one must take care not to assume that simple Yukawa exchange is the only dynamical effect present.

The proposal of charmed-meson molecules (such as the $J^{PC} = 1^{++}$ $S$-wave pair $\bar D^0 D^{*0}$ plus its charge-conjugate) actually far predates the discovery of $X(3872)$: In fact, such states were first proposed~\cite{Voloshin:1976ap} almost immediately after the discovery of charmed mesons!  And since $\Delta E \equiv m_{X(3872)} - m_{D^0} - m_{D^{*0}} = -40 \pm 90$~keV, how could $X(3872)$ {\em not\/} be a $\bar D^0 D^{*0}$ molecule (even though $\Delta E$ is so tiny, casting doubt on a model based exclusively upon Yukawa exchange)?

Several of the other exotics, both tetraquarks and pentaquarks, lie just below di-hadron thresholds: as just two examples, $m_{X(3915)} - m_{D_s} - m_{D_s} \cong -15$~MeV and $m_{P_c (4312)} - m_{\Sigma_c^+} - m_{D^0} \cong -7$~MeV\@.  But {\em not all\/} of them do!  In particular, some lie just {\em above\/} di-hadron thresholds, such as $m_{Z_c^+ (3900)} - m_{D^+} - m_{D^{*0}} \cong 10$~MeV and $m_{Z_b^+ (10610)} -m_{B^0} - m_{B^{*+}} \cong +3$~MeV\@.  Moreover, we have noted that several exotic candidates are nowhere near obvious thresholds; and in addition, not every di-hadron threshold exhibits a strong enhancement that might be interpreted as an exotic state.

The minuscule binding energy of $X(3872)$ also suggests through the de~Broglie relation a characteristic size of $\hbar / \sqrt{2\mu \Delta E} > 10$~fm, which is actually larger than almost all nuclei.  And yet, lots of $X(3872)$ are produced promptly ({\it i.e.}, at the initial $c\bar c$ production point) in collider experiments.  Such a result suggests that $X(3872)$ may contain both molecular and compact, strongly bound components.  Fortunately, a criterion for the compositeness of a state was developed by Weinberg long ago~\cite{Weinberg:1965zz} (originally to study the deuteron): From production lineshape parameters such as the scattering length, one may extract a parameter $Z$ that measures compositeness (which is 0 for a molecule and 1 for a compact, elementary state).  The latest data from LHCb~\cite{LHCb:2020xds} yields $Z > 0.15$ for $X(3872)$, indicating a state that is largely but not entirely molecular.

\subsection{Threshold/Rescattering/Cusp Effects}

The presence of nearby di-hadron thresholds has other important influences on how certain exotic states are interpreted.  They appear through complex-analytic features of amplitudes such as poles, branch points, and Riemann sheets, and represent the revival of an extensive 1960s technology that formed a pillar of strong-interaction studies in the pre-QCD era.  The simplest example is the opening of an on-shell di-hadron channel, which creates a branch point (as a function of Mandelstam $s$ of the pair) most apparent in the imaginary part of the amplitude $\Pi$.  For an $S$-wave hadron pair, the optical theorem causes Im~$\Pi$ to grow as a square-root function of $s$ proportional to phase space.  Feeding Im~$\Pi$ back into Cauchy’s theorem creates a {\em cusp\/} in Re $\Pi$ that resembles a resonance.  If an intrinsic state with its own pole occurs anywhere nearby, then the observed pole can be ``dragged'' toward threshold~\cite{Bugg:2008wu}, providing a mechanism for the occurrence of near-threshold states.

Significant threshold effects can also appear through exchanges of virtual particles between the threshold pair, which are called {\em rescattering effects}, and can be especially prominent if some of the internal particle lines approach their mass shell.  Prime examples include so-called {\em triangle singularities}, which arise when the resonance candidate couples to the di-hadron state through a triangular loop diagram.  The results of such calculations (first applied to exotics by Ref.~\cite{Chen:2011pv}) create new analytic structures that change production lineshapes in ways that can be probed by experiment.  In particular, rescattering diagrams can generate poles, either below or above threshold [like $Z_c (3900)$!].  Depending upon where poles occur in the various Riemann sheets for the amplitude, they are called {\em bound states}, {\em virtual states}, or {\em resonances} (see Ref.~\cite{Guo:2017jvc}); all of these phenomena are genuine physical entities that have been observed in nuclear scattering experiments.

It is important to note that Weinberg’s $Z$ can be applied to any scattering states, including the examples discussed here.  Thus, a state containing a large component of hadrons $A$ \& $B$ by this measure is still called an $AB$ molecule in modern terminology, independent of any Yukawa-type interaction between them.  In this context, LHCb showed~\cite{LHCb:2020xds} that $X(3872)$ is mostly a $\bar D^0 D^{*0}$ molecule, even though traditional $\pi$-exchange does not provide a useful description.  Likewise, analysis by the JPAC Collaboration~\cite{Pilloni:2016obd} concludes that $Z_c (3900)$ is simultaneously a virtual state above the $\bar D D^*$ threshold and a molecule.

\subsection{Hadrocharmonium}

The hadrocharmonium (or more generally, hadroquarkonium) model was originally developed~\cite{Dubynskiy:2008mq} to explain why some exotics, such as $Y(4260)$ [later resolved into $Y(4220)$ and other states], $Y(4660)$, and $Z_c (4430)$, prefer to decay to specific charmonium states ($J/\psi$, $\psi^\prime$, $\chi_c$, or $h_c$), rather than to open-charm $D^{(*)}$ meson pairs.  In this model, the exotic state possesses a quarkonium core in a specific spin state and with a specific radial excitation [{\it e.g.}, $s_{c\bar c} = 1$ and $n=2$ for $Z_c(4430)$ leads to a preferential decay to $\psi^\prime$, as is experimentally observed].  However, some states like $Y(4220)$ have subsequently been seen to decay into both $J/\psi (s_{c\bar c} = 1)$ and $h_c (s_{c\bar c} = 0)$ channels, and into open-charm pairs ($\pi^+ D^0 D^{*-}$) as well.  Consequently, hadrocharmonium is no longer as prominent in theoretical studies; nevertheless, it has left a lasting impact in exotics studies because the use of the heavy-quark spin basis (and approximate heavy-quark spin symmetry) remains quite effective for understanding spin structure and decays in molecular and diquark states.

\subsection{Diquark-Antidiquark Hadrons}

An important but often neglected feature of SU(3)$_{\rm color}$ is the attraction between a diquark pair $qq$ in the channel ${\bf 3} \otimes {\bf 3} \to \bar {\bf 3}$.  At short distances (where single-gluon exchange dominates), the coupling is {\em fully half as strong\/} as the $q\bar q$ attraction ${\bf 3} \otimes \bar {\bf 3} \to {\bf 1}$ that forms the short-distance limit of the confining interaction.  This ``Casimir scaling'' is derived at leading order in QCD, but it actually survives up to 3$^{\rm rd}$ order~\cite{Anzai:2010td}.  Meanwhile, the existence of the diquark attraction out to distances as large as $\sim 1$~fm is supported by lattice calculations~\cite{Bali:2000un}.

Diquarks have long been analyzed as quasiparticles within baryons (particularly $\Lambda$), even before QCD~\cite{Lichtenberg:1967zz}.  By the late 1970s, peculiar features of scalar mesons like $a_0 (980)$, $f_0 (980)$ led to the proposal that they are tetraquarks consisting of a diquark-antidiquark ($\delta^{\vphantom\dagger}$-$\bar \delta$) pair~\cite{Jaffe:1976ig}.

The diquark paradigm was first applied to heavy-quark exotics in Ref.~\cite{Maiani:2004vq}, when rather few candidates were known; it quickly became apparent that different spin states of the model assigned to $X(3872)$ and $Z_c (3900)$ occur with wrong multiplicities and mass ordering.  However, the model also admitted a clever fix with a much more successful phenomenology~\cite{Maiani:2014aja}:  The dominant spin couplings in the tetraquark state are taken to be only those {\em within\/} each diquark, which is to be expected if $\delta$ and $\bar \delta$ become somewhat separated from each other within the hadron.

This separation is manifested in the {\em dynamical diquark picture}~\cite{Brodsky:2014xia}: In high-energy $B$-decay or collider processes, $\delta$ and $\bar \delta$ can form and separate before color recombination into mesons occurs, allowing identification of the state as $\delta^{\vphantom\dagger}$-$\bar \delta$.  The color flux tube that forms between the $\delta^{\vphantom\dagger}$-$\bar \delta$ pair converts their kinetic energy into potential energy, and brings them essentially to rest before they can decay to mesons; thus, one can describe the state using the Born-Oppenheimer approximation, which allows for the creation of a predictive model.  This picture supports the formation of exotics that are spatially large but still strongly coupled, because the diquarks are colored objects.  It can also be extended to depict pentaquarks as diquark-{\em triquark} $\left[ \bar c_{\bar 3} (ud)_{\bar 3} \right]_{\bf 3}$ systems~\cite{Lebed:2015tna}, using consecutive color-{\bf 3} attractions to form triquark quasiparticles.

With the introduction of isospin dependence between $\delta^{\vphantom\dagger}$-$\bar \delta$ pair, the model produces a full spectrum of states ({\it e.g.}, all 12 states in the ground-state $c\bar c q \bar q^\prime$ multiplet)~\cite{Giron:2019cfc}, and it has been applied to every flavor sector.  Most recently, the dynamical diquark model has successfully been generalized to incorporate threshold effects (the {\em diabatic formalism}) to provide a natural explanation for why exotics like $X(3872)$ lie near di-hadron thresholds~\cite{Lebed:2022vks,Lebed:2023kbm}.

\section{Lattice QCD}

Since the resolution to the question of which theoretical picture most accurately describes exotics remains ambiguous, why not just put the 4 (or 5) quarks on the lattice?  The short answer is that lattice-QCD techniques have not yet mastered a number of complications that occur for the exotic candidates.  To begin with, the original method for lattice simulations cannot handle unstable, above-threshold resonances.  But newer ones can (assuming the presence of only one dominant decay channel): One may extract quantities like decay widths by studying the scattering matrix in a finite box, the so-called {\em L\"{u}scher formalism}~\cite{Luscher:1991cf}.  On the other hand, simulations for exotics with more than one decay channel or with dominant 3-body decays are not yet mature.  Furthermore, lattice calculations are tricky for problems for which several energy or length scales occur, such as for systems with heavy constituents but also small binding energies; indeed, the best simulations of a state as well known as the deuteron~\cite{NPLQCD:2011naw} still cannot definitively explain why its binding energy is as small as 2.2~MeV\@.

With these limitations in mind, what has been accomplished so far to describe exotics on the lattice?  A sampling of some prominent results includes:
\begin{itemize}

\item A state like $X(3872)$ appears, but its presence requires both $c\bar c$ and $\bar D D^*$ (but not diquark) components, and it has no charged partner~\cite{Padmanath:2015era}.

\item Evidence for the existence of $Z_c (3900)$ is mixed; one group says it appears through coupling of $\bar D D^*$ and $J/\psi \pi$ channels~\cite{HALQCD:2016ofq}, while another group, using L\"{u}scher formalism, has not found it yet~\cite{CLQCD:2019npr}.

\item The observed hidden-bottom $Z_b (10610)$ is found (correctly) to lie just above the $\bar B B^*$ threshold~\cite{Prelovsek:2019ywc}.

\item $P_c$ pentaquarks are now seen in simulations  that include couplings to $\Sigma_c \bar D$ and $\Sigma_c \bar D^*$ channels~\cite{Xing:2022ijm}.

\item A variety of lattice simulations, starting with Ref.~\cite{Bicudo:2015vta}, predict that the lowest $bb\bar u \bar d$ tetraquark is actually {\em stable\/} against strong decay.

\end{itemize}

\section{Summary and Prospects}

What have we learned in the past two decades, and what are we likely to learn in the next few years?  Most obviously, tetraquarks and pentaquarks exist in large numbers, and it is growing increasingly clear that no single theoretical paradigm explains all $> 60$ of the heavy-quark exotics.  Moreover, many additional such states are likely to be discovered in the near future.

Threshold effects seem to be essential for a realistic description of many of the exotic candidates, but even those seem to be insufficient if one allows for only the single di-hadron component corresponding to the threshold [like $\bar D^0 D^{*0}$ for $X(3872)$].  Furthermore, not every di-hadron threshold produces a strong signal for an exotic candidate.

Diquark models (analogously to quarkonium models) produce specific spectra that can be verified or falsified for each observed state, and these models can be systematically improved to include threshold effects.

Heavy-quark spin symmetry (which is most cleanly probed in the decays of exotics to conventional quarkonium states) provides important clues on underlying spin structure of exotics.  More generally,
transitions between exotic states [like $Y(4220) \to Z_c (3900)$], and branching ratios of exotics to quarkonium and to heavy open-flavor pairs, will be essential in determining which exotics have similar underlying structures.

\begin{acknowledgments}
This work was supported by the National Science Foundation (NSF) under 
Grants No.\ PHY-1803912 and PHY-2110278.
\end{acknowledgments}

\bibliographystyle{JHEP}
\bibliography{diquark}

\end{document}